\input{hyperlatex}
\documentstyle[preprint,eqsecnum,aps]{revtex}

\newcommand{\f}[2]{\frac{#1}{#2}}
\newcommand{\s}[1]{\sqrt{#1}}

\begin{document}

\preprint{\vbox{
\hbox{UCSD/PTH 96--01}
\hbox{hep-ph/96mmnnn}
}}
\title{SU(3) Decomposition of Two-Body $B$ Decay Amplitudes}
\author{Benjam\'{\i}n Grinstein and Richard F. Lebed}
\address{Department of Physics, University of California at
San Diego, La Jolla, CA 92093}
\date{February 1, 1996}
\maketitle
\begin{abstract}
	We present the complete flavor SU(3) decomposition of decay
amplitudes for decays of the triplet ($B^+_u$, $B^0_d$, $B^0_s$) of
$B$ mesons nonleptonically into two pseudoscalar mesons.  This
analysis holds for arbitrarily broken SU(3) and can be used to
generate amplitude relations when physical arguments permit one to
neglect or relate any of the reduced amplitudes.
\end{abstract}



\narrowtext

\section{Introduction}
	The current understanding of charge-changing quark transitions
in terms of the Cabibbo-Kobayashi-Maskawa (CKM) mixing
matrix\cite{CKM} is becoming progressively more open to scrutiny.
Large numbers of new experimental results involving the physics of
$b$-quarks permit one to perform incisive tests on the ill-known
parameters of the third quark generation.  The decays of $B$ mesons
should provide an ample testing ground for determining quantities of
interest.  Whether the CKM matrix is really unitary, the significance of
the size hierarchy observed in CKM elements, and the origin of CP
violation are issues that might be resolved, at least in part,
within the next decade, owing to improved experimental information.

	Before this optimistic program can be undertaken, however, one
requires tools of analysis that facilitate the extraction of the
relevant parameters from the data.  The chief problem is that
virtually all decays of $B$ mesons contain at least one hadron, and
fundamental quark-related quantities are notoriously difficult to
extract from the corresponding hadronic quantities.  Furthermore, the
neutral $B$ mesons $B^0_d$ and $B^0_s$ mix with their antiparticles,
thus complicating particle identification.  Nevertheless, one can make
progress in both of these areas from knowing the symmetry of the
underlying theory and a few of its dynamical properties.

	A case in point is an interesting series of papers\cite{GHLR},
in which it is claimed that one can disentangle CKM elements and
strong-interaction final-state phase shifts from nonleptonic two-body
$B$ decays.  The key ingredient in this analysis is the idea\cite{SWZ}
that the numerous possible experimental measurements of nonleptonic
decays of the mesons $B^+_u$, $B^0_d$, and $B^0_s$ can be related
using the flavor SU(3) approximate symmetry of the strong interaction
Lagrangian.  Such a symmetry exists owing to the relative smallness of
the $u$, $d$, and $s$ quark masses compared to the QCD scale
$\Lambda_{\rm QCD}$.  According to Refs.~\cite{GHLR}, with $i)$ a
large number of these rates eventually measured, and $ii)$ mild
dynamical assumptions on the strong interaction physics interpreted at
the quark level, one obtains through group theory an over-determined
system of equations, which can be solved to isolate CKM elements and
strong-interaction final-state phase shifts.  The $B$ decays in this
approach are described in terms of naive quark diagrams, some of which
are taken to be suppressed on physical grounds; for instance, the
diagram describing the annihilation of the valence quark-antiquark
pair is said to be suppressed by a factor of the meson decay constant
over its mass.  Then, since the SU(3) flavor structure of the
intermediate quark lines is simple, one can calculate a set of decay
amplitudes in terms of reduced SU(3) amplitudes and look for relations
between them.  Because the coefficients are proportional to CKM
elements and strong interaction phases, solving the system of
equations permits one to extract these quantities.  However, the quark
diagram approach has two major drawbacks.  First, the exact nature of
SU(3) group theory is not fully manifest in such a description, so
that relations thus derived tend to appear as surprising cancellations
between diagrams; and second, the dynamical assumptions mentioned
above are at best only semi-quantitative and do not lend themselves
well to systematic corrections.

	In this paper, we remedy the first problem by presenting the
complete SU(3) decomposition for two-body nonleptonic decays of the
$B$ triplet to pseudoscalar mesons with and without charm, including
arbitrary breaking of SU(3) (as well as isospin) symmetry.  Many of
the relations derived in Refs.~\cite{GHLR} correspond to the
suppression of Hamiltonian operators transforming under particular
irreducible representations of SU(3), while others correspond to
chance cancellations owing to the phenomenological neglect of
particular quark diagrams in their description.  The second problem
will be addressed in a future publication~\cite{GLfut}.  The current
paper is a reference work that permits one to perform the SU(3)
analysis of these decays using any particular set of dynamical
assumptions.  Its chief advantage is that, while one can write down an
Hamiltonian with an arbitrary number of parameters to produce a model,
the number of SU(3) reduced amplitudes for given initial and final
states is a finite and exactly calculable number, and all relations
obtained due to symmetry alone are made explicit.

	This paper is organized as follows: In Sec.\ 2, we describe
two equivalent means by which one may obtain the necessary
Clebsch-Gordan coefficients to decompose the physical amplitudes in
terms of SU(3) amplitudes.  In Sec.\ 3, we explain the counting of
these two types of amplitudes in fully broken SU(3).  In Sec.\ 4 the
means by which relations between physical amplitudes are obtained is
explored.  We consider examples arising from the assumption of an
unbroken SU(3) Hamiltonian defined through a four-quark operator, as
well as the inclusion of linear SU(3) breaking.  Sec.\ 5 discusses
directions for future work and concludes.  The group-theoretical
results are contained in the Appendix.

\section{SU(3) Group Theory}
	A full treatment of the SU(3) decomposition of physical
amplitudes is completely equivalent to the application of the
Wigner-Eckart theorem for the group SU(3).  One obtains the amplitudes
for the decays of physical particles into reduced SU(3) amplitudes,
and the connection between these two bases are simply Clebsch-Gordan
coefficients.  There are two ways to achieve such a decomposition.

	The first method is to work with roots, weights, and ladder
operators in the usual manner of Wigner to obtain the desired
coefficients.  Tables of SU(3) Clebsch-Gordan coefficients for smaller
irreducible representations have existed for some time\cite{deS},
although tables containing all the representations one requires can be
more difficult to find\cite{Kaed}.  Even with the coefficients in
hand, one must convolve several layers of Clebsch-Gordan coefficients
to complete this task; this follows because combining two
representations is a binary operation, and each additional initial-
and final-state particle requires another product.  One must also take
care to observe the proper phase conventions, which ultimately arise
when one requires representations and their conjugates to obey
simultaneously the same phase convention for ladder operators.

	The second method is to work directly with tensors.  Indeed,
Clebsch-Gordan coefficients are simply the coefficients of the
couplings of tensors that have been appropriately symmetrized,
normalized, and rendered traceless.  The appeal of this approach is
that one can immediately think of the tensors as pieces of the
interaction Hamiltonian.  In any case, both methods must give
identical results, and we have confirmed this through direct
calculation.

	In either approach, the phase differences between
representations and their conjugates must be included in some fashion.
These phases arise from the convention one adopts in relating physical
states to weights in group space.  The most convenient means of doing
so is to take the fundamental and fundamental conjugate
representations to consist of the quark flavor states
\begin{equation}
{\bf 3} = \left(
\begin{array}{r}
u \\ d \\ s
\end{array}
\right) \, , \hspace{3em}
\bar {\bf 3} = \left(
\begin{array}{r}
\bar d \\ - \bar u \\ \bar s
\end{array}
\right) .
\end{equation}
Including an additional sign for each $\bar u$ permits one to assign
the physical mesons to the weights in SU(3) representations without
additional phases.  In this convention, the flavor wavefunctions of
the mesons of interest are:
\begin{eqnarray}
& & K^+ = +u \bar s, \hspace{2em} K^0 = +d \bar s, \label{first} \\
& & \pi^+ = +u
\bar d, \hspace{2em} \pi^0 = -\f{1}{\s{2}} (+u \bar u - d \bar d),
\hspace{2em} \pi^- = -d \bar u, \\
& & \eta_8 = -\f{1}{\s{6}} (+ u \bar u + d \bar d - 2 s \bar s), \\
& & \bar K^0 = +s \bar d, \hspace {2em} K^- = -s \bar u, \\
& & \eta_1 = -\f{1}{\s{3}} (+u \bar u +d \bar d +s \bar s), \label{last}
\end{eqnarray}
for the light meson nonet $P$,
\begin{eqnarray}
& & B^+_u = +\bar b u, \hspace{2em} B^0_d = +\bar b d, \hspace{2em}
B^0_s = +\bar b s,
\end{eqnarray}
for the triplet $B$'s,
\begin{eqnarray}
& & \bar D^0 = +\bar c u, \hspace{2em} D^- = +\bar c d, \hspace{2em}
D_s^- = +\bar c s,
\end{eqnarray}
for the triplet $D$'s, and
\begin{eqnarray}
& & D^0 = -c \bar u, \hspace{2em} D^+ = +c \bar d, \hspace{2em} D_s^+
= +c \bar s,
\end{eqnarray}
for the antitriplet $D$'s.  In addition, there is the charmonium
singlet
\begin{equation}
\eta_c = +\bar c c.
\end{equation}
The physical mesons $\eta$, $\eta'$ are defined through the SO(2)
rotation
\begin{equation} \label{mix}
\left( \begin{array}{c}
\eta \\ \eta'
\end{array} \right) =
\left( \begin{array}{rr}
\mbox{} -\cos \theta & \mbox{} +\sin \theta \\ \mbox{} -\sin \theta &
\mbox{} -\cos \theta
\end{array} \right) \left( \begin{array}{c}
\eta_8 \\ \eta_1
\end{array} \right),
\end{equation}
where the peculiar sign conventions on the mixing are defined so that
the angle $\theta$ agrees with that of Gilman and Kauffmann\cite{GK},
in which the mixing is phenomenologically determined to assume a value
of $\theta \simeq -20^{\circ}$.

	In the tensor approach the signs from the above phase
convention are ignored.  Indeed, the light meson pseudoscalar octet is
taken to be represented by the traceless matrix
\begin{equation}
M = \left( \begin{array}{ccc}
\frac{1}{\sqrt{2}} \pi^0 + \frac{1}{\sqrt{6}} \eta_8 & \pi^+ & K^+ \\
\pi^- & -\frac{1}{\sqrt{2}} \pi^0 + \frac{1}{\sqrt{6}} \eta_8 & K^0 \\
K^- & \bar K^0 & -\sqrt{\frac{2}{3}} \eta_8
\end{array} \right) ,
\end{equation}
so that the signs of $\pi^0$, $\pi^-$, $\eta_8$, and $K^-$ are
opposite to those in Eqs.~(\ref{first})--(\ref{last}).  Similarly, in
a direct tensor approach one drops the above signs for both $\eta_1$
and $D^0$.  These differences in convention result in different signs
in the physical amplitudes; there is one relative sign change for each
time one of the aforementioned particles appears in an amplitude.
These differences are trivial to implement.

\section{Counting Amplitudes}
	In completely broken flavor SU(3) for a sector of given
quantum numbers (corresponding to a Hamiltonian with particular
eigenvalues of the diagonal generators taken to be the isospin third
component $I_3$ and the hypercharge $Y$) there must be exactly as many
reduced SU(3) amplitudes as distinct physical processes.  This is just
a statement of the completeness of the amplitude basis in either
physical or group-theoretical terms.  For example, consider the case
of $B \rightarrow PP$, where $P$ here and throughout the paper
designates the light pseudoscalar nonet consisting of $\pi$, $K$,
$\eta$, $\eta'$ (the octet and singlet components of the physical
$\eta$, $\eta'$ are designated $\eta_8$ and $\eta_1$, respectively,
and are related by Eq.~(\ref{mix})).  Because group theory relates
processes with the same Hamiltonian $\Delta I_3$ and $\Delta Y$ (which
are equivalent to electric charge $\Delta Q$ and strangeness $\Delta
S$ changes\footnote{The exact relations are $Q = I_3 + \f{1}{2} Y +
Q_{\rm h}$ and $S = Y -\f{1}{3} T$, where $Q_{\rm h}$ is the charge of
quarks not belonging to the SU(3) flavor triplet, and $T$ is the
triality of the SU(3) representation, which is the number of
fundamental representation indices modulo 3 required to build a tensor
transforming under the given representation.  For ${\bf 3}$,$\bar {\bf
3}$, this number is $\pm 1$, whereas for octets and singlets it is
zero.}), and in any process electric charge is conserved, sets of
amplitudes with a particular strangeness change $\Delta S$ are related
by group theory.  For processes not changing strangeness between
initial and final states, for example, one counts twelve amplitudes
when both $P$'s are octet mesons, four with one octet $P$ and the
other being the singlet $\eta_1$, and one ($B^0_d \rightarrow \eta_1
\eta_1$) with both $P$'s being singlets.  It must be that there are
exactly equal numbers of SU(3) reduced amplitudes for each set of
particle representations, and this is indeed the case (see Appendix).

	In order to count these amplitudes, one must construct the
most general possible transformation structure for the interaction
Hamiltonian.  Because the Hamiltonian connects initial to final states
via the matrix elements $\left< f | {\cal H} | i \right>$, the most
general interaction Hamiltonian consists of exactly those
representations ${\bf R}$ appearing in ${\bf f} \otimes \bar {\bf i}$.
The labels $i$ and $f$ here are used to denote both states and SU(3)
representations.  There is one further complication in that states of
SU(3) representations are uniquely distinguished when eigenvalues of
not only $I_3$ and $Y$ but also the isospin Casimir $I^2$ are
specified.  The full reduced SU(3) amplitude is thus described by the
notation $\left< f || R_I || i \right>$.

	If the two final-state particles are both in the same SU(3)
representation ${\bf f}'$, then their amplitudes obey one further
restriction owing to the Pauli Exclusion Principle.  Because the
initial and both final particles are spin-zero, the spatial part of
the final wavefunction is $s$-wave and therefore symmetric under
interchange of particle labels.  But because the final-state bosons
transform under the same representation of SU(3), they are identical
particles modulo SU(3) indices, and the total wavefunction must be
symmetric under exchanging the two particles.  Therefore, the flavor
wavefunction alone must also be symmetric under this interchange.
This symmetrization, which we write as ${\bf f} = ({\bf f}' \otimes
{\bf f}')_S$, eliminates a number of possible representations.  For
example, in the case of $B \rightarrow PP$ with $P$ in the octet, the
amplitudes {\it a priori\/} transform as
\begin{equation}
{\bf 8} \otimes {\bf 8} = {\bf 1} \oplus {\bf 8}_S \oplus {\bf 8}_A
\oplus {\bf 10} \oplus \overline{\bf 10} \oplus {\bf 27} ,
\end{equation}
but the only ones allowed by the Exclusion Principle are
\begin{equation}
({\bf 8} \otimes {\bf 8})_S = {\bf 1} \oplus {\bf 8}_S \oplus {\bf
27}.
\end{equation}
Note that this restriction fails to hold if the final state does not
possess a completely symmetric spatial wavefunction, as is the case
for arbitrary initial and final spin states.  It is also required that
the final state particles occupy two weights in the same
representation ${\bf f}'$, not merely two distinct copies of ${\bf
f}'$.

	It may seem odd that what we call the reduced matrix element
$\left< f || R_I || i \right>$ is dependent upon the isospin Casimir
$I$, not just the SU(3) irreducible representation, of the Hamiltonian
operator.  This seems to contradict the Wigner-Eckart theorem, which
states that {\it all\/} of the matrix elements of a particular tensor
operator, for states in given initial and final state representations,
are related by Clebsch-Gordan coefficients.  While the theorem is
certainly true for each tensor operator contributing to a physical
process, the Hamiltonian itself may have dynamical coefficients that
are unequal for different components of a given representation.  To be
explicit, let us adopt the notation for SU(3) Clebsch-Gordan
coefficients of de Swart\cite{deS}.  The coefficient coupling the
representations ${\bf R_a} \otimes {\bf R_b} \rightarrow {\bf R_c}$ is
indicated by
\begin{equation}
\left( \begin{array}{ccc}
R_a & R_b & R_c \\
I_a I_{a3} Y_a & I_b I_{b3} Y_b & I_c I_{c3} Y_c
\end{array} \right) .
\end{equation}
For brevity, let us denote the quantum numbers $I,I_3,Y$ within an
SU(3) representation by the collective label $\nu$.  Then the physical
amplitude ${\cal A}$ is decomposed into our reduced matrix elements by
\begin{equation} \label{amp}
{\cal A} (i^{R_c}_{\nu_c} \rightarrow f^{R_a}_{\nu_a}
f^{R_b}_{\nu_b}) =  (-1)^{\left(I_3+\frac{Y}{2} + \frac{T}{3}
\right)_{\overline{R}_c}}
\sum_{\stackrel{\scriptstyle R', \nu'}{R, \nu}}
\left( \begin{array}{ccc}
R_a & R_b & R' \\ \nu_a & \nu_b & \nu'
\end{array} \right)
\left( \begin{array}{ccc}
R' & \overline{R}_c & R \\ \nu' & -\nu_c & \nu
\end{array} \right)
\left< R' || R_\nu || R_c \right> .
\end{equation}
Note the order of coupling of the representations: First, the
final-state representations are coupled via ${\bf R_a} \otimes {\bf
R_b} \rightarrow {\bf R'}$, and this representation in turn is coupled
to the Hamiltonian through the conjugate of the initial
representation, ${\bf R'} \otimes \overline{\bf R}_{\bf c} \rightarrow
{\bf R}$, or ${\bf f} \otimes \overline{\bf i} \rightarrow {\cal H}$.
As pointed out in the Appendix, coupling in this order ensures that
the Clebsch-Gordan matrices are orthogonal.  The phase in the above
expression arises from the fact that we use not the initial
representation ${\bf i}$, but its conjugate $\overline{\bf i}$; $T$ is
the triality of the representation $\overline{\bf R}_{\bf c}$, as
defined in Sec.\ 2, and guarantees the reality of the phase.  In the
present case, it induces an additional sign on decays of $B^+_u$ but
not $B^0_d$ or $B^0_s$.  On the other hand, we may choose to couple
representations in a more standard order.  If we decompose the
Hamiltonian as
\begin{equation}
{\cal H} = \sum_{R, \nu} c_{R, \nu} {\cal H}^R_\nu ,
\end{equation}
then the expression for the physical amplitude becomes
\begin{equation}
{\cal A} (i^{R_c}_{\nu_c} \rightarrow f^{R_a}_{\nu_a}
f^{R_b}_{\nu_b}) =  \sum_{R, \nu} c_{R, \nu}
\sum_{R', \nu'}
\left( \begin{array}{ccc}
R_a & R_b & R' \\ \nu_a & \nu_b & \nu'
\end{array} \right)
\left( \begin{array}{ccc}
R & R_c & R' \\ \nu & \nu_c & \nu'
\end{array} \right)
\left< R' || R || R_c \right> ,
\end{equation}
which produces the usual Wigner-Eckart reduced amplitude $\left< R' ||
R || R_c \right>$, independent of any other quantum numbers, but the
Hamiltonian coefficients $c$ have nontrivial $\nu$ dependence in
general.  Note that the products performed here are ${\bf R_a} \otimes
{\bf R_b} \rightarrow {\bf R'}$ and ${\bf R} \otimes {\bf R_c}
\rightarrow {\bf R'}$, or ${\cal H} \otimes {\bf i} \rightarrow {\bf
f}$.  The relation between the two reduced amplitudes can be
established by the use of symmetry and completeness relations
satisfied by the SU(3) Clebsch-Gordan coefficients~\cite{deS}.  Up to
a phase dependent upon the representations coupled,
\begin{equation}
\left< R' || R_\nu || R_c \right> = c_{R, \nu} \sqrt{\frac{{\rm dim} \;
R'}{{\rm dim} \; R}} \left< R' || R || R_c \right> .
\end{equation}
From this expression we see that the two definitions differ in that
ours absorbs the dynamical coefficient appearing with the given
operator in the Hamiltonian.  Once one specifies, as in the Appendix,
the values of $I_3$ and $Y$ for the Hamiltonian operator, the only
free index remaining in $\nu$ is $I$.  Finally, in the case that the
two final-state particles are in the same representation as described
above so that $R_a = R_b$, then in the above expressions one makes the
symmetrizing substitution
\begin{equation} \label{sym}
\left( \begin{array}{ccc}
R_a & R_b & R' \\ \nu_a & \nu_b & \nu'
\end{array} \right) \rightarrow \f{1}{\sqrt{2}} \left[
\left( \begin{array}{ccc}
R_a & R_b & R' \\ \nu_a & \nu_b & \nu'
\end{array} \right) +
\left( \begin{array}{ccc}
R_b & R_a & R' \\ \nu_b & \nu_a & \nu'
\end{array} \right) \right] .
\end{equation}
Note that, for identical final-state particles ($\nu_a = \nu_b$), this
substitution induces an additional factor of $\sqrt{2}$ in the
amplitude.  While this factor complicates what we mean by the physical
amplitude, the substitution Eq.~(\ref{sym}) nevertheless preserves the
orthogonality of the Clebsch--Gordan matrices, as can be shown by
either symbolic or direct numerical calculation.  A trivial example of
this factor is illustrated by the analogous case of spin SU(2), where
starting with the four two-spin basis states $\uparrow \uparrow$,
$\uparrow \downarrow$, $\downarrow \uparrow$, $\downarrow \downarrow$,
the symmetrization indicated by Eq.~(\ref{sym}) correctly gives
$\frac{1}{\sqrt{2}} ( \uparrow \downarrow + \downarrow \uparrow )$,
but also $\frac{1}{\sqrt{2}} (2 \uparrow \uparrow )$ and
$\frac{1}{\sqrt{2}} (2 \downarrow \downarrow )$.

	We close this section by explaining the physical
interpretation of the amplitudes computed in the Appendix.  Whereas
the physical amplitude for distinct spinless final-state particles
$(\nu_a \neq \nu_b)$ simply squares to the measurable rate, one must
include additional Bose symmetry factors when the final-state
particles are identical ($R_a = R_b$ and $\nu_a = \nu_b$).  We have
noted that the amplitude given by Eq.~(\ref{amp}) with the
substitution Eq.~(\ref{sym}) for identical particles is a factor
$\sqrt{2}$ larger than the naive definition of the physical amplitude.
On the other hand, to obtain the decay rate one multiplies the naive
amplitude by an exchange factor 2! (giving the usual physical
amplitude), squares, and divides by an identical particle factor 2! in
the rate to avoid multiple counting.  It follows that the rate for
$\nu_a = \nu_b$ is twice the naive amplitude squared, or simply the
amplitude of Eq.~(\ref{amp}) squared, the same as for $\nu_a \neq
\nu_b$.  Because the universal rule $rate = amplitude^2$ is simpler
than keeping track of factors of two in certain cases, we present in
the Appendix the amplitudes given by Eq.~(\ref{amp}), using
Eq.~(\ref{sym}) in all cases where $R_a = R_b$.

\section{Amplitude Relations}
	The above counting describes how one enumerates the complete
set of amplitudes for arbitrarily broken SU(3).  This counting holds
even is there is no good physical reason to organize particles into
SU(3) multiplets.  For example, one could take eight arbitrary
particles and call them an octet of SU(3), and the group-theoretical
decomposition, which is purely mathematical, would remain true.
Clearly one requires physical input to make practical use of the group
theory.  If one finds a physical reason why a particular SU(3) reduced
amplitude should vanish, then the particular combination of physical
amplitudes to which it is equal also vanishes.  This corresponds to
taking the inverse of the transformation matrix ${\cal O}$, for which
each row decomposes a particular physical amplitude into reduced SU(3)
amplitudes.  However, as noted in the Appendix, the SU(3) reduced
matrix elements are normalized so that the basis transformation
matrices ${\cal O}$ are orthogonal, hence ${\cal O}^{-1} = {\cal
O}^T$, and the relation associated with the vanishing of a particular
SU(3) reduced amplitude is obtained by merely reading off the entries
from the corresponding column of ${\cal O}$.  The relations also hold
for the charge-conjugated states, although the presence of CP
violation means that amplitudes for individual processes do not
necessarily equal the amplitudes for their conjugate processes.
Finally, as discussed in Sec.\ 2, these matrices are obtained using a
particular phase convention.  Choosing another either results in
changing the signs of particular particle states in terms of their
quark-antiquark indices or changing the signs in the definition of
reduced matrix elements.  These correspond respectively to changing
the signs of rows or columns of the matrices in the Appendix,
operations which do not affect their orthogonality.

	One begins by assuming a form for the unbroken
Hamiltonian.  For the case of $B$ decay, this is of course the
four-quark Hamiltonian derived from the tree-level weak interaction
\begin{equation}
{\cal H}_{\rm int} = \f{4G_F}{\s{2}} V_{{q_1} b}^* V_{{q_2} {q_3}}
(\bar b_L \gamma^\mu {q_1}_L) (\overline{q_2}_L \gamma_\mu {q_3}_L) ,
\end{equation}
where $q_{1,2}$ are charge +2/3 ($u$,$c$) quarks and $q_3$ is a charge
--1/3 ($d$,$s$) quark.  Note that there are several physical
assumptions already included in this ansatz.  In writing the
Hamiltonian this way, the physical $B$ decay is assumed to be
dominated by the decay of the $\bar b$ quark into a $\bar c$ or $\bar
u$ quark with the emission of a virtual $W^-$, which subsequently
decays into a quark-antiquark pair.  All other contributions involving
QCD renormalization effects or penguins, for example, are considered
negligible in this limit.

	One can analyze the Hamiltonian for each case of flavor
content.  The field operators $q_i$, $\bar q_i$ for $q_i = u,d,s$
transform as components of $\bar {\bf 3}$, ${\bf 3}$ respectively,
because these operators respectively destroy initial-state quarks and
antiquarks; $q_i = c$ is of course a singlet in flavor SU(3).  For the
case $\Delta C = 0$ with no $c \bar c$ pair in the final state ($B
\rightarrow PP$), the SU(3) representations allowed by ${\cal H}_{\rm
int}$ are those in
\begin{equation} \label{reps}
\bar {\bf 3} \otimes \bar {\bf 3} \otimes {\bf 3} = \bar {\bf 3}
\oplus \bar {\bf 3} \oplus {\bf 6} \oplus \overline{\bf 15} .
\end{equation}
The redundancy of the $\bar {\bf 3}$ representation in the Hamiltonian
is irrelevant, because one cannot distinguish in this case the two
contributions, which transform in the same way.  It follows that the
lowest-order Hamiltonian has pieces transforming as $\bar {\bf 3}$,
${\bf 6}$, and $\overline{\bf 15}$, but not ${\bf 24}$ or
$\overline{\bf 42}$ (see Eqs.~(\ref{s088}), (\ref{s188})), and so for
either $\Delta S = 0$ or $\Delta S = +1$ there are five amplitude
relations corresponding to the five vanishing amplitude combinations
transforming under ${\bf 24}$ or $\overline{\bf 42}$ in
Eqs.~(\ref{s088}), (\ref{s188}).  It is readily seen why the ${\bf 24}$
and $\overline{\bf 42}$ representations do not appear at leading
order: These representations require three indices in either the
fundamental or fundamental conjugate representation (see
Eq.~(\ref{young})), and this is impossible with a four-quark
(two-quark, two-antiquark) operator.  These relations may be obtained
as the amplitude combinations that obtain no contributions from the
operators analyzed in Refs.~\cite{SWZ}.

	In order to analyze isospin content, we must further specify
the number of strange quarks created or destroyed in the process.  For
$B \rightarrow PP$ with $\Delta S = 0$, the possible isospins are
those in $\f{1}{2} \otimes \f{1}{2} \otimes \f{1}{2}$, namely
$I=\f{1}{2}, \f{3}{2}$; and for $\Delta S = +1$ only two quarks are
light, so $I=0,1$ are possible.  In the SU(3) symmetry limit, this
does not eliminate any additional amplitudes besides the ones
mentioned above.  The analysis for other values of $\Delta S$ in $B
\rightarrow PP$ is straightforward: More $s$ or $\bar s$ quarks
implies that the maximum allowed $I$ from a four-quark Hamiltonian is
smaller.

	Let us now consider SU(3)-breaking corrections to the
lowest-order Hamiltonian.  The simplest such breaking originates
through insertions of the strange quark mass,
\begin{equation} \label{brk}
{\cal H}_s = m_s \bar s s ,
\end{equation}
which transforms as an $I=0$, $Y=0$ octet plus singlet in SU(3).
Clearly neither piece changes the isospin of the Hamiltonian; this
would be accomplished by insertions of the up or down masses, which
are much smaller.  Let us consider SU(3) breaking linear in $m_s$.  In
the case of $B \rightarrow PP$, the Hamiltonian contains pieces
transforming under
\begin{equation}
(\bar {\bf 3} \oplus {\bf 6} \oplus \overline{\bf 15}) \otimes ({\bf
1} \oplus {\bf 8}) = \bar {\bf 3} \oplus {\bf 6} \oplus \overline{\bf
15} \oplus {\bf 24} \oplus \overline{\bf 42} ,
\end{equation}
not counting multiplicities.  Comparing to Eqs.~(\ref{s088}) and
(\ref{s188}), we see that every allowed Hamiltonian SU(3)
representation occurs.  Nothing is gained group-theoretically by
stopping at linear order in strange-quark masses.  This interesting
conclusion turns out to be true for every decay considered in the
Appendix, when all allowed values of $\Delta S$ are considered.  It is
generally true for an amplitude with a total of three mesons (each of
which is group-theoretically a quark-antiquark state) between the
initial and final states, because the Hamiltonian is a six-quark
(three-quark, three-antiquark) operator, and so every representation
that can connect the initial and final states can occur in the
Hamiltonian.  On the other hand, because $\bar s s$ is an $I=0$
operator, the restriction that only $I=\f{1}{2}, \f{3}{2}$ are allowed
for $\Delta S = 0$ and $I=0, 1$ are allowed for $\Delta S = +1$
remains true even when the lowest-order Hamiltonian is corrected with
an arbitrary number of $\bar s s$ insertions.  One then still has the
relations
\begin{eqnarray}
\left< 27 || \overline{42}_{I=\f{5}{2}} || 3 \right> & = & 0 \; \;
({\rm for} \; \Delta S = 0), \\
\left< 27 || \overline{42}_{I=2} || 3 \right> & = & 0 \; \; ({\rm for
} \; \Delta S = +1).
\end{eqnarray}
These relations follow entirely from the fact that we have broken
the SU(3) symmetry in the Hamiltonian, but not isospin.

	A similar analysis holds for Hamiltonians with different
flavor contents.  In general, the group theory is simpler for decays
with charm quarks in the final state, since charm transforms as an
SU(3) singlet.

	Amplitude relations may also appear for particular values of
$\Delta S$ since entire SU(3) representations may be disallowed by the
particular Hamiltonian used.  For example, in the case $B \rightarrow
D \bar D$ or $B \rightarrow \eta_c P$ with $\Delta S = +1$
(Eqs.~(\ref{bdds1}), (\ref{etac})), the lowest-order tree-level
Hamiltonian is
\begin{equation} \label{hcc}
{\cal H}_0 \sim (\bar b c) (\bar c s) ,
\end{equation}
where we have suppressed all except flavor indices.  In terms of
flavor SU(3), this operator falls into the $\bar {\bf 3}$
representation.  If we include one insertion of SU(3) breaking of the
form Eq.~(\ref{brk}), then
\begin{equation} \label{hcc2}
\delta {\cal H} \sim (\bar b c) (\bar c s) (\bar s s) .
\end{equation}
Group-theoretically, one may first combine $(s \otimes s) \in \bar
{\bf 6} \oplus {\bf 3}$.  However, ${\bf 3}$ does not contain a state
with the hypercharge $Y = +2/3$ of $(s \otimes s)$ and so does not
occur.  Taking the product of these representations with the remaining
$\bar s$,
\begin{equation}
\bar {\bf 6} \otimes {\bf 3} = \overline{\bf 15} \oplus \bar {\bf 3},
\hspace{2em} {\bf 3} \otimes {\bf 3} = {\bf 6} \oplus \bar {\bf 3} .
\end{equation}
In particular, a Hamiltonian ${\bf 6}$, which is allowed if all values
of $\Delta S$ are considered, does not occur in the case $\Delta S =
+1$, and so one obtains an amplitude relation,
\begin{equation}
\left< 8 || 6_{I=1} || 3 \right> = 0.
\end{equation}
Similar reasoning applies to the decays $B \rightarrow \bar D P$ with
$\Delta S = +1$ (Eq.~(\ref{bdps1})), for which one finds
\begin{equation}
\left< 15 || 10_{I=\frac{3}{2}} || 3 \right> = 0 ,
\end{equation}
for tree-level plus first-order SU(3) symmetry-breaking terms in the
Hamiltonian.  Note that this conclusion is specific to the choice of
form for the Hamiltonian.  It should also be noted that these
relations may be obtained through the isospin analysis described
above.

	In greatest generality, the SU(3) reduced matrix elements are
all independent, because the physical amplitudes need not be related
in any way.  Each such reduced matrix element corresponds to a
component of the Hamiltonian transforming under a particular
representation of SU(3) with a particular value of isospin, and so in
the most general situation the coefficient of each component is
independent.  However, in the usual case, one uses a Hamiltonian with
particular operators (typically written in terms of quark fields) that
can be explicitly decomposed under SU(3).  Then the reduced matrix
elements allowed by these operators are related by a product of the
Clebsch-Gordan coefficients obtained by projecting the given operators
onto SU(3) representations multiplied by the explicit coefficients of
the original operators themselves.  As an example, consider the $B
\rightarrow PP$ reduced matrix elements $\left< f ||
\overline{15}_{I=1} || i \right>$ (in $\Delta S = 0$) and $\left< f ||
\overline{15}_{I=\f{3}{2}} || i \right>$ (in $\Delta S = +1$).  A
tree-level analysis suggests that these matrix elements are dominated
by the Hamiltonian
\begin{equation} \label{tree}
{\cal H}_{\rm int} = \f{4G_F}{\s{2}} \left[ V_{ub}^* V_{ud}
\left( \bar b_L \gamma^\mu u_L \right) \left( \bar u_L \gamma_\mu d_L
\right) + V_{ub}^* V_{us}
\left( \bar b_L \gamma^\mu u_L \right) \left( \bar u_L \gamma_\mu s_L
\right) \right] .
\end{equation}
The gluonic penguin diagram does not contribute to the part of the
Hamiltonian transforming as a $\overline{\bf 15}$ when SU(3) is
unbroken since the gluon is an SU(3) singlet, and so too must be the
quark-antiquark pair produced by it.  Thus the penguin process has
flavor content only through the decays $\bar b \rightarrow \bar d$ ($I
= \f{1}{2}$) or $\bar b \rightarrow \bar s$ ($I = 0$), which transform
as components of a $\bar {\bf 3}$.  The SU(3) structure of the
four-quark operators in Eq.~(\ref{tree}) is manifest, and one can
immediately project them onto the $\overline{\bf 15}$ to obtain the
corresponding Clebsch-Gordan coefficients.  The two Hamiltonian
operators otherwise differ only in their coefficients, two
combinations of CKM elements.  The result of the calculation is
\begin{equation} \label{rat1}
\f{\left< f || \overline{15}_{I=1} || i \right>}{\left< f ||
\overline{15}_{I=\f{3}{2}} || i \right>} = +\f{\s{3}}{2}
\f{V_{us}}{V_{ud}},
\end{equation}
regardless of the initial- or final-state representations ($i$,$f$) of
the particles.  The corresponding expression within a given
strangeness sector is even simpler, because then the CKM element for
each $\overline{\bf 15}$ component is the same.  In particular,
\begin{equation} \label{rat2}
\f{\left< f || \overline{15}_{I=\f{1}{2}} || i \right>}{\left< f ||
\overline{15}_{I=\f{3}{2}} || i \right>} = +\f{1}{2\s{2}},
\hspace{3em} \f{\left< f || \overline{15}_{I=0} || i \right>}{\left< f
|| \overline{15}_{I=1} || i \right>} = +\f{1}{\s{2}},
\end{equation}
for $\Delta S = 0$ and $\Delta S = +1$, respectively.
The key to the simplicity of the ratios in this example is that only
one operator structure (the four-quark operator) dominates the
Hamiltonian for the given decays; when several different operator
structures contribute to the Hamiltonian for a particular decay,
ratios like Eqs.~(\ref{rat1}), (\ref{rat2}) are replaced by
\begin{equation} \label{rat3}
\f{\left< f || R_I || i \right>}{\left< f || R_{I'}' || i \right>} =
\f{\sum_j c_j \, {\cal C}_j}{\sum_k c_k \, {\cal C}_k'} ,
\end{equation}
where ${\cal C}$ and ${\cal C}'$ are Clebsch-Gordan coefficients, and
$c$ are coefficients of the different components of the Hamiltonian.
This is the case when, for example, penguin diagrams, diagrams
involving the participation of the spectator quark, or SU(3)
corrections are significant.  For example, in the case of the
$\overline{\bf 15}$, ${\cal O}(m_s)$ corrections to the Hamiltonian
(\ref{tree}) introduce an additional operator transforming as a
$\overline{\bf 15}$, since in Eq.~(\ref{reps}) $\overline{\bf 15}
\otimes {\bf 8} \supset \overline{\bf 15}$.  On the other hand, even
lowest-order Hamiltonian operators transforming under $\bar{\bf 3}$ or
${\bf 6}$ corrected by the SU(3) octet breaking produce $\overline{\bf
15}$'s.  Then the relations (\ref{rat1}), (\ref{rat2}) are replaced
with ones of the form (\ref{rat3}).  For the $\overline{\bf 15}$ such
corrections may be small, but it is often the case that two or more
operators of the same numerical order and distinct coefficients
appear when considering a particular Hamiltonian representation; in
such cases, Eq.~(\ref{rat3}) must be used.

	We comment briefly on the relations that can be derived
through the above analysis and those derived in Refs.~\cite{GHLR}.
Each relation unbroken by the full set of quark interactions described
in \cite{GHLR} appears because the quark diagrams may be
re-interpreted as collections of quark fields with definite SU(3)
flavor properties in a Hamiltonian, and as such give rise to a set of
SU(3) irreducible representations.  In this way, the quark diagram
approach is group-theoretically equivalent to the more formal
approach, as was first pointed out by Zeppenfeld\cite{SWZ}.  All
possible representations that do not appear in the Hamiltonian thus
give rise to amplitude relations.  The analysis of isospin relations
is particularly straightforward, as we have discussed.  Indeed, in
addition to the numerous isospin relations derived in \cite{GHLR}, we
add one they have omitted,
\begin{equation}
{\cal A} (B_s^0 \rightarrow \eta_c \pi^0) = 0,
\end{equation}
which vanishes because neither the Hamiltonian Eq.~(\ref{hcc}) nor the
SU(3)-breaking correction (\ref{hcc2}) contains an $I=1$ piece, while
the process in question is pure $I=1$ (only the pion carries isospin
in the decay).  The latter fact is corroborated by a quick glance at
the third row of Eq.~(\ref{etac}).  The SU(3) relations in \cite{GHLR}
may be obtained through analysis like that leading to
Eqs.~(\ref{rat1})--(\ref{rat2}), whereas their more detailed analysis
of neglecting certain quark diagrams would be obtained in our language
by decomposing the corresponding Hamiltonian quark operator into a
combination of different SU(3) representations.  Finally, a number of
$\eta , \eta'$ relations are possible when one includes a value for
the mixing of $\eta_1, \eta_8$ as discussed in Sec.\ 2.

\section{Prospects and Improvements}

	The purpose of the discussion and formulas contained in this
work is to provide a complete analysis of the group-theoretical
problem of the decay of $B$ mesons into two pseudoscalar mesons.
Because it is completely general in the mathematical sense, it
provides a valuable tool of analysis for researchers performing
computations of such processes.  But this strength is also its
weakness: No physical content was included, except for trivial
illustrative examples.  Nevertheless, once any particular Hamiltonian
is adopted, one can immediately see exactly which physical amplitudes
vanish or are related for symmetry reasons, and why.

	The fully consistent application of this analysis to the
physical problem of unambiguously extracting CKM elements and
strong-interaction final-state phase shifts requires one to impose an
SU(3) decomposition on the full Hamiltonian taking part in the $B$
decay, including short-distance QCD corrections and SU(3) symmetry
breaking.  Because gluons transform as singlets under flavor SU(3),
many extremely complicated diagrams involving large numbers of gluons
and sea quark-antiquark pairs can still be taken into account in a
simple way using the flavor symmetry.  On the other hand,
complications may arise, such as the presence of potentially important
electroweak penguin diagrams\cite{DHO}, which have a nontrivial flavor
structure and thus must be treated carefully.  By choosing reduced
matrix elements insensitive to these corrections, one may be able to
avoid this difficulty.  Amplitude relations that survive these
corrections will be invaluable for studying the detailed structure of
the CKM matrix.

	There is much to be learned even from the amplitudes that are
not suppressed.  In this case, the interesting question is whether the
reduced matrix elements obey the numerical hierarchy predicted by a
more naive analysis, based upon some physical model, that estimates
the size of operator coefficients.  With enough rates eventually
measured, one will be able to compute reduced matrix elements
directly, without recourse to model-dependent assumptions.  For
example, the issue of whether gluonic penguins or tree-level
amplitudes dominate a given process will be directly resolved,
inasmuch as their corresponding Hamiltonians transform differently
under SU(3).  This knowledge can be applied to many other interactions
involving heavy quarks.  We plan to address these questions in greater
detail in a future publication.

\vskip1.2cm
{\it Acknowledgments}
\hfil\break
This work is supported in part by the Department of Energy under
contract DOE-FG03-90ER40546.  The research of B. G. is also funded in
part by the Alfred P. Sloan Foundation.

\appendix
\section*{Clebsch-Gordan Tables}
	Presented below is the complete decomposition in terms of
SU(3) reduced amplitudes for decays of a flavor-triplet $B$ meson into
a pair of pseudoscalars with $\Delta C = 0$, $+1$, or $-1$.  The phase
convention is chosen to agree with that of Condon and
Shortley\cite{CS} for isospin SU(2), which is defined by two
conditions.  First, the phase in the definition of the isospin raising
and lowering operators acting on a given isospin eigenstate is chosen
to be +1; this establishes phases within a particular isomultiplet.
Second, to establish the relative phase between multiplets (in this
case isomultiplets with a common value of hypercharge), one considers
the couplings of the two factor representations ($I^{(a)}$, $I^{(b)}$,
with $I^{(a)} \ge I^{(b)}$) to a given product representation $I$.
Then, for the state of highest weight in the product multiplet ($I_3 =
I$), the coupling
\begin{displaymath}
\left< I^{(a)} \, I^{(a)}_3 ; \, I^{(b)} \, I^{(b)}_3 | I \, I \right>
\end{displaymath}
is chosen to have phase $+1$ when $I^{(a)}_3$ is the largest such
value that a nonzero coupling occurs.  The relative phases for states
with different values of hypercharge are fixed in the SU(3) convention
of de Swart\cite{deS}, in that the first condition of the
Condon-Shortley convention is extended to hold for both isospin and
$V$-spin ($V_3 \equiv \f{3}{4} Y + \f{1}{2} I_3$).  The arbitrary
choice of $V$-spin instead of $U$-spin or some combination leads to
the convention in Sec.\ 2 that the fundamental conjugate state $\bar
u$ transforms with phase opposite to that of $\bar d$, $\bar s$.  The
phase convention on the physical states is also described in Sec.\ 2.
Each matrix, as one can check, is orthogonal, thus establishing that
the SU(3) reduced amplitudes form an orthonormal basis equivalent to
the amplitudes ${\cal A}$.

	The physical decay rate is obtained in all cases by squaring
the quantity ${\cal A}$.  If the two final-state particles are
identical, the quantity ${\cal A}$ is the usual physical amplitude
(which already includes an identical particle factor) divided by
$\sqrt{2}$, as described in Sec.\ 3.

	All SU(3) representations are indicated below merely by their
dimensions; this creates no problem for us, since no two distinct
representations with the same dimensionality appear in this analysis.
For convenience, we also present here the equivalent weight notation
$(p,q)$, where $p$ and $q$ respectively indicate the number of
fundamental and fundamental conjugate indices in the tensor
representation.  The Young tableau then consists of a row of $p+q$
boxes over a row of $q$ boxes.  The representation is labeled with a
bar if $p<q$.
\begin{equation} \label{young}
\begin{array}{rrr}
{\bf 1} = (0,0), & {\bf 3} = (1,0), & {\bf 6} = (2,0), \\
{\bf 8} = (1,1), & {\bf 10} = (3,0), & {\bf 15} = (2,1), \\
{\bf 24} = (3,1), & {\bf 27} = (2,2), & {\bf 42} = (3,2) .
\end{array}
\end{equation}

	The expressions below are divided into sections first by
particle content of the final state, and then by the number of units
of strangeness changed in the $B$ decay.

\subsection{$B \rightarrow P,P$}
\subsubsection{\rm{$\Delta S = 0$ ($\Delta I_3 = +\f{1}{2}$, $\Delta Y =
-\f{1}{3}$)}}
For $P,P$ both octet mesons, denote ${\bf u}_{8,8}^{S=0} = {\cal
O}_{8,8}^{S=0} \, {\bf v}_{8,8}^{S=0}$, where
\begin{equation} \label{s088}
{\bf u}_{8,8}^{S=0} =
\left( \begin{array}{c}
{\cal A} (B^+_u \rightarrow K^+ \bar K^0) \\
{\cal A} (B^+_u \rightarrow \pi^+ \pi^0) \\
{\cal A} (B^+_u \rightarrow \pi^+ \eta_8) \\
{\cal A} (B^0_d \rightarrow K^+ K^-) \\
{\cal A} (B^0_d \rightarrow \pi^+ \pi^-) \\
{\cal A} (B^0_d \rightarrow \pi^0 \pi^0) \\
{\cal A} (B^0_d \rightarrow \pi^0 \eta_8) \\
{\cal A} (B^0_d \rightarrow \eta_8 \eta_8) \\
{\cal A} (B^0_d \rightarrow K^0 \bar K^0) \\
{\cal A} (B^0_s \rightarrow \bar K^0 \pi^0) \\
{\cal A} (B^0_s \rightarrow \bar K^0 \eta_8) \\
{\cal A} (B^0_s \rightarrow K^- \pi^+)
\end{array} \right) , \hspace{5em}
{\bf v}_{8,8}^{S=0} =
\left( \begin{array}{c}
\left< 1 || \bar 3_{I=\frac 1 2} || 3 \right> \\
\left< 8 || \bar 3_{I=\frac 1 2} || 3 \right> \\
\left< 8 || 6_{I=\frac 1 2} || 3 \right> \\
\left< 8 || \overline{15}_{I=\frac 1 2} || 3 \right> \\
\left< 8 || \overline{15}_{I=\frac 3 2} || 3 \right> \\
\left< 27 || \overline{15}_{I=\frac 1 2} || 3 \right> \\
\left< 27 || \overline{15}_{I=\frac 3 2} || 3 \right> \\
\left< 27 || 24_{I=\frac 1 2} || 3 \right> \\
\left< 27 || 24_{I=\frac 3 2} || 3 \right> \\
\left< 27 || \overline{42}_{I=\frac 1 2} || 3 \right> \\
\left< 27 || \overline{42}_{I=\frac 3 2} || 3 \right> \\
\left< 27 || \overline{42}_{I=\frac 5 2} || 3 \right>
\end{array} \right) ,
\end{equation}
and ${\cal O}_{8,8}^{S=0} =$
\begin{displaymath}
\hspace{-1.6em}
\left( \begin{array}{cccccccccccc}
0 & +\f{3}{2\s{10}} & +\f{1}{2}\s{\f{3}{5}} & +\f{1}{2\s{10}} &
+\f{1}{\s{5}} & -\f{4}{3\s{15}} & -\f{1}{3\s{30}} & -\f{4}{3\s{15}} &
+\f{1}{3\s{3}} & -\f{2}{3\s{15}} & -\f{1}{3}\s{\f{5}{6}} & 0 \\
0 & 0 & 0 & 0 & 0 & 0 & -\f{1}{2}\s{\f{5}{3}} & 0 & -\f{1}{\s{6}} & 0 &
-\f{1}{2\s{15}} & -\s{\f{2}{5}} \\
0 & -\f{1}{2}\s{\f{3}{5}} & -\f{1}{\s{10}} & -\f{1}{2\s{15}} &
-\s{\f{2}{15}} & -\f{2}{3}\s{\f{2}{5}} & -\f{1}{6\s{5}} &
-\f{2}{3}\s{\f{2}{5}} & +\f{1}{3\s{2}} & -\f{1}{3}{\s{\f{2}{5}}} &
-\f{\s{5}}{6} & 0 \\
+\f{1}{2} & +\f{1}{\s{10}} & 0 & +\f{1}{\s{10}} & -\f{1}{\s{5}} &
-\f{1}{6\s{15}} & +\f{1}{3\s{30}} & -\f{1}{3}\s{\f{5}{3}} &
-\f{1}{3\s{3}} & +\f{2}{3\s{15}} & +\f{1}{3}\s{\f{5}{6}} & 0 \\
+\f{1}{2} & -\f{1}{2\s{10}} & +\f{1}{2}\s{\f{3}{5}} & -\f{3}{2\s{10}}
& 0 & -\f{1}{6\s{15}} & -\f{1}{3}\s{\f{5}{6}} & +\f{1}{3\s{15}} &
-\f{1}{3\s{3}} & -\f{1}{3\s{15}} & -\f{1}{3\s{30}} & +\f{1}{\s{5}} \\
-\f{1}{2\s{2}} & +\f{1}{4\s{5}} & -\f{1}{2}\s{\f{3}{10}} &
+\f{3}{4\s{5}} & 0 & +\f{1}{6\s{30}} & -\f{1}{3}\s{\f{5}{3}} &
-\f{1}{3\s{30}}& -\f{1}{3}\s{\f{2}{3}} & +\f{1}{3\s{30}} &
-\f{1}{3\s{15}} & +\s{\f{2}{5}} \\
0 & -\f{1}{2}\s{\f{3}{10}} & -\f{1}{2\s{5}} & -\f{1}{2\s{30}} &
+\f{2}{\s{15}} & -\f{2}{3\s{5}} & +\f{1}{3\s{10}} & -\f{2}{3\s{5}} &
-\f{1}{3} & -\f{1}{3\s{5}} & +\f{1}{3}\s{\f{5}{2}} & 0 \\
-\f{1}{2\s{2}} & -\f{1}{4\s{5}} & +\f{1}{2}\s{\f{3}{10}} &
-\f{3}{4\s{5}} & 0 & +\f{1}{2}\s{\f{3}{10}} & 0 & -\s{\f{3}{10}} & 0 &
+\s{\f{3}{10}} & 0 & 0 \\
-\f{1}{2} & +\f{1}{2\s{10}} & +\f{1}{2}\s{\f{3}{5}} & -\f{1}{2\s{10}}
& -\f{1}{\s{5}} & -\f{7}{6\s{15}} & +\f{1}{3\s{30}} & +\f{1}{3\s{15}}
& -\f{1}{3\s{3}} & -\f{4}{3\s{15}} & +\f{1}{3}\s{\f{5}{6}} & 0 \\
0 & +\f{3}{4\s{5}} & -\f{1}{2}\s{\f{3}{10}} & -\f{3}{4\s{5}} & 0 &
+\f{1}{3}\s{\f{2}{15}} & -\f{1}{3}\s{\f{5}{3}} & -\f{1}{3\s{30}} &
+\f{2}{3}\s{\f{2}{3}} & -\f{1}{3}\s{\f{2}{15}} & +\f{1}{3}\s{\f{5}{3}}
& 0 \\
0 & +\f{1}{4}\s{\f{3}{5}} & -\f{1}{2\s{10}} & -\f{1}{4}\s{\f{3}{5}} &
0 & -\s{\f{2}{5}} & 0 & +\f{1}{\s{10}} & 0 & +\s{\f{2}{5}} & 0 & 0 \\
0 & -\f{3}{2\s{10}} & +\f{1}{2}\s{\f{3}{5}} & +\f{3}{2\s{10}} & 0 &
-\f{2}{3\s{15}} & -\f{1}{3}\s{\f{5}{6}} & +\f{1}{3\s{15}} &
+\f{2}{3\s{3}} & +\f{2}{3\s{15}} & +\f{1}{3}\s{\f{5}{6}} & 0
\end{array} \right)  \! ,
\end{displaymath}
\begin{equation} \end{equation}
\begin{equation}
\left(
\begin{array}{c}
{\cal A} (B^+_u \rightarrow \pi^+ \eta_1) \\
{\cal A} (B^0_d \rightarrow \pi^0 \eta_1) \\
{\cal A} (B^0_d \rightarrow \pi^+ \eta_1) \\
{\cal A} (B^0_s \rightarrow \pi^0 \eta_1)
\end{array}
\right) = 
\left( \begin{array}{cccc}
-\f{1}{2}\s{\f{3}{2}} & -\f{1}{2} & -\f{1}{2\s{6}} & -\f{1}{\s{3}} \\
-\f{\s{3}}{4} & -\f{1}{2\s{2}} & -\f{1}{4\s{3}} & +\s{\f{2}{3}} \\
+\f{1}{4} & -\f{1}{2}\s{\f{3}{2}} & +\f{3}{4} & 0 \\
-\f{1}{2}\s{\f{3}{2}} & +\f{1}{2} & +\f{1}{2}\s{\f{3}{2}} & 0
\end{array} \right)
\left(
\begin{array}{c}
\left< 8 || \bar 3_{I=\f{1}{2}} || 3 \right> \\
\left< 8 || \bar 6_{I=\f{1}{2}} || 3 \right> \\
\left< 8 || \overline{15}_{I=\f{1}{2}} || 3 \right> \\
\left< 8 || \overline{15}_{I=\f{3}{2}} || 3 \right> \\
\end{array} \right),
\end{equation}
and
\begin{equation}
{\cal A}(B^0_d \rightarrow \eta_1 \eta_1) = (+1) \left< 1 || \bar
3_{I=\f{1}{2}} || 3 \right> .
\end{equation}
\subsubsection{\rm{$\Delta S = +1$ ($\Delta I_3 = 0$, $\Delta Y =
+\f{2}{3}$)}}
For $P,P$ both octet mesons, denote ${\bf u}_{8,8}^{S=1} = {\cal
O}_{8,8}^{S=1} \, {\bf v}_{8,8}^{S=1}$, where
\begin{equation} \label{s188}
{\bf u}_{8,8}^{S=1} =
\left( \begin{array}{c}
{\cal A} (B^+_u \rightarrow K^0 \pi^+) \\
{\cal A} (B^+_u \rightarrow K^+ \pi^0) \\
{\cal A} (B^+_u \rightarrow K^+ \eta_8) \\
{\cal A} (B^0_d \rightarrow K^+ \pi^-) \\
{\cal A} (B^0_d \rightarrow K^0 \pi^0) \\
{\cal A} (B^0_d \rightarrow K^0 \eta_8) \\
{\cal A} (B^0_s \rightarrow \pi^+ \pi^-) \\
{\cal A} (B^0_s \rightarrow \pi^0 \pi^0) \\
{\cal A} (B^0_s \rightarrow \pi^0 \eta_8) \\
{\cal A} (B^0_s \rightarrow \eta_8 \eta_8) \\
{\cal A} (B^0_s \rightarrow K^+ K^-) \\
{\cal A} (B^0_s \rightarrow K^0 \bar K^0)
\end{array} \right) , \hspace{5em}
{\bf v}_{8,8}^{S=1} =
\left( \begin{array}{c}
\left< 1 || \bar 3_{I=0} || 3 \right> \\
\left< 8 || \bar 3_{I=0} || 3 \right> \\
\left< 8 || 6_{I=1} || 3 \right> \\
\left< 8 || \overline{15}_{I=0} || 3 \right> \\
\left< 8 || \overline{15}_{I=1} || 3 \right> \\
\left< 27 || \overline{15}_{I=0} || 3 \right> \\
\left< 27 || \overline{15}_{I=1} || 3 \right> \\
\left< 27 || 24_{I=1} || 3 \right> \\
\left< 27 || 24_{I=2} || 3 \right> \\
\left< 27 || \overline{42}_{I=0} || 3 \right> \\
\left< 27 || \overline{42}_{I=1} || 3 \right> \\
\left< 27 || \overline{42}_{I=2} || 3 \right>
\end{array} \right) ,
\end{equation}
and ${\cal O}_{8,8}^{S=1} =$
\begin{displaymath}
\hspace{-1.6em}
\left( \begin{array}{cccccccccccc}
0 & +\f{3}{2\s{10}} & -\f{1}{2}\s{\f{3}{5}} & +\f{1}{2}\s{\f{3}{10}} &
+\f{1}{2}\s{\f{3}{5}} & +\f{1}{3\s{5}} & -\f{1}{\s{10}} &
-\f{1}{\s{10}} & +\f{1}{3\s{2}} & +\f{1}{3\s{10}} & 0 & -\f{1}{3} \\
0 & -\f{3}{4\s{5}} & +\f{1}{2}\s{\f{3}{10}} & -\f{1}{4}\s{\f{3}{5}} &
-\f{1}{2}\s{\f{3}{10}} & -\f{1}{3\s{10}} & -\f{7}{6\s{5}} &
-\f{1}{3\s{5}} & +\f{1}{3} & -\f{1}{6\s{5}} & -\f{1}{3\s{2}} &
-\f{\s{2}}{3} \\
0 & +\f{1}{4}\s{\f{3}{5}} & -\f{1}{2\s{10}} & +\f{1}{4\s{5}} &
+\f{1}{2\s{10}} & -\s{\f{3}{10}} & -\f{1}{2\s{15}} & +\f{2}{\s{15}} &
0 & -\f{1}{2}\s{\f{3}{5}} & -\f{1}{\s{6}} & 0 \\
0 & -\f{3}{2\s{10}} & -\f{1}{2}\s{\f{3}{5}} & -\f{1}{2}\s{\f{3}{10}} &
+\f{1}{2}\s{\f{3}{5}} & -\f{1}{3\s{5}} & -\f{1}{\s{10}} &
-\f{1}{\s{10}} & -\f{1}{3\s{2}} & -\f{1}{3\s{10}} & 0 & +\f{1}{3} \\
0 & +\f{3}{4\s{5}} & +\f{1}{2}\s{\f{3}{10}} & +\f{1}{4}\s{\f{3}{5}} &
-\f{1}{2}\s{\f{3}{10}} & +\f{1}{3\s{10}} & -\f{7}{6\s{5}} &
-\f{1}{3\s{5}} & -\f{1}{3} & +\f{1}{6\s{5}} & -\f{1}{3\s{2}} &
+\f{\s{2}}{3} \\
0 & +\f{1}{4}\s{\f{3}{5}} & +\f{1}{2\s{10}} & +\f{1}{4\s{5}} &
-\f{1}{2\s{10}} & -\s{\f{3}{10}} & +\f{1}{2\s{15}} & -\f{2}{\s{15}} &
0 & -\f{1}{2}\s{\f{3}{5}} & +\f{1}{\s{6}} & 0 \\
+\f{1}{2} & +\f{1}{\s{10}} & 0 & -\s{\f{3}{10}} & 0 & +\f{1}{6\s{5}} &
0 & 0 & +\f{\s{2}}{3} & -\f{1}{3\s{10}} & 0 & +\f{1}{3} \\
-\f{1}{2\s{2}} & -\f{1}{2\s{5}} & 0 & +\f{1}{2}\s{\f{3}{5}} & 0 &
-\f{1}{6\s{10}} & 0 & 0 & +\f{2}{3} & +\f{1}{6\s{5}} & 0 & +\f{\s{2}}{3} \\
0 & 0 & +\f{1}{\s{5}} & 0 & +\f{1}{\s{5}} & 0 & -\s{\f{2}{15}} &
+\s{\f{2}{15}} & 0 & 0 & +\f{1}{\s{3}} & 0 \\
-\f{1}{2\s{2}} & +\f{1}{2\s{5}} & 0 & -\f{1}{2}\s{\f{3}{5}} & 0 &
-\f{3}{2\s{10}} & 0 & 0 & 0 & +\f{3}{2\s{5}} & 0 & 0 \\
+\f{1}{2} & -\f{1}{2\s{10}} & -\f{1}{2}\s{\f{3}{5}} &
+\f{1}{2}\s{\f{3}{10}} & -\f{1}{2}\s{\f{3}{5}} & -\f{1}{2\s{5}} &
-\f{1}{3}\s{\f{2}{5}} & +\f{1}{3}\s{\f{2}{5}} & 0 & +\f{1}{\s{10}} &
+\f{1}{3} & 0 \\
-\f{1}{2} & +\f{1}{2\s{10}} & -\f{1}{2}\s{\f{3}{5}} &
-\f{1}{2}\s{\f{3}{10}} & -\f{1}{2}\s{\f{3}{5}} & +\f{1}{2\s{5}} &
-\f{1}{3}\s{\f{2}{5}} & +\f{1}{3}\s{\f{2}{5}} & 0 & -\f{1}{\s{10}} &
+\f{1}{3} & 0 
\end{array} \right) \! ,
\end{displaymath}
\begin{equation} \end{equation}
\begin{equation}
\left(
\begin{array}{c}
{\cal A} (B^+_u \rightarrow K^+ \eta_1) \\
{\cal A} (B^0_d \rightarrow K^0 \eta_1) \\
{\cal A} (B^0_s \rightarrow \pi^0 \eta_1) \\
{\cal A} (B^0_s \rightarrow \eta_8 \eta_1)
\end{array}
\right) = 
\left( \begin{array}{cccc}
-\f{1}{2}\s{\f{3}{2}} & +\f{1}{2} & -\f{1}{2\s{2}} & -\f{1}{2} \\
-\f{1}{2}\s{\f{3}{2}} & -\f{1}{2} & -\f{1}{2\s{2}} & +\f{1}{2} \\
0 & +\f{1}{\s{2}} & 0 & +\f{1}{\s{2}} \\
-\f{1}{2} & 0 & +\f{\s{3}}{2} & 0
\end{array} \right)
\left(
\begin{array}{c}
\left< 8 || \bar 3_{I=0} || 3 \right> \\
\left< 8 || \bar 6_{I=1} || 3 \right> \\
\left< 8 || \overline{15}_{I=0} || 3 \right> \\
\left< 8 || \overline{15}_{I=1} || 3 \right> \\
\end{array} \right),
\end{equation}
and
\begin{equation}
{\cal A}(B^0_s \rightarrow \eta_1 \eta_1) = (+1) \left< 1 || \bar
3_{I=0} || 3 \right> .
\end{equation}
\subsubsection{\rm{$\Delta S = -1$ ($\Delta I_3 = +1$, $\Delta
Y=-\f{4}{3}$)}}
\begin{equation}
\left(
\begin{array}{c}
{\cal A} (B^+_u \rightarrow \bar K^0 \pi^+) \\
{\cal A} (B^0_d \rightarrow \bar K^0 \pi^0) \\
{\cal A} (B^0_d \rightarrow \bar K^0 \eta_8) \\
{\cal A} (B^0_d \rightarrow K^- \pi^+) \\
{\cal A} (B^0_s \rightarrow \bar K^0 \bar K^0)
\end{array}
\right) = 
\left( \begin{array}{ccccc}
0 & -\f{1}{3}\s{\f{5}{2}} & -\f{2}{3} & -\f{1}{6} & -\f{1}{2} \\
-\s{\f{3}{10}} & -\f{2}{3\s{5}} & -\f{1}{3\s{2}} & -\f{1}{3\s{2}}
& +\f{1}{\s{2}} \\
-\f{1}{\s{10}} & +\f{1}{\s{15}} & -\f{1}{\s{6}} & +\s{\f{2}{3}} & 0 \\
+\s{\f{3}{5}} & -\f{1}{3\s{10}} & -\f{1}{3} & +\f{1}{6} & +\f{1}{2} \\
0 & -\f{\s{5}}{3} & +\f{\s{2}}{3} & +\f{\s{2}}{3} & 0
\end{array} \right)
\left(
\begin{array}{c}
\left< 8 || \overline{15}_{I=1} || 3 \right> \\
\left< 27 || \overline{15}_{I=1} || 3 \right> \\
\left< 27 || 24_{I=1} || 3 \right> \\
\left< 27 || \overline{42}_{I=1} || 3 \right> \\
\left< 27 || \overline{42}_{I=2} || 3 \right> \\
\end{array} \right),
\end{equation}
and
\begin{equation}
{\cal A}(B^0_d \rightarrow \bar K^0 \eta_1) = (+1) \left< 1 ||
\overline{15}_{I=1} || 3 \right> .
\end{equation}
\subsubsection{\rm{$\Delta S = +2$ ($\Delta I_3 = -\f{1}{2}$, $\Delta
Y=+\f{5}{3}$)}}
\begin{equation}
\left(
\begin{array}{c}
{\cal A} (B^+_u \rightarrow K^+ K^0) \\
{\cal A} (B^0_d \rightarrow K^0 K^0) \\
{\cal A} (B^0_s \rightarrow K^0 \pi^0) \\
{\cal A} (B^0_d \rightarrow K^+ \pi^-) \\
{\cal A} (B^0_s \rightarrow K^0 \eta_8)
\end{array}
\right) = \left(
\begin{array}{ccccc}
0 & -\f{1}{3}\s{\f{5}{2}} & +\f{2}{3} & -\f{1}{3\s{2}} & -\f{\s{2}}{3}
\\
0 & -\f{\s{5}}{3} & -\f{\s{2}}{3} & -\f{1}{3} & +\f{1}{3} \\
-\s{\f{3}{10}} & +\f{1}{6\s{5}} & +\f{\s{2}}{3} & -\f{1}{6} &
+\f{2}{3} \\
+\s{\f{3}{5}} & -\f{1}{3\s{10}} & +\f{1}{3} & +\f{1}{3\s{2}} &
+\f{\s{2}}{3} \\
-\f{1}{\s{10}} & -\f{1}{2}\s{\f{3}{5}} & 0 & +\f{\s{3}}{2} & 0
\end{array} \right)
\left(
\begin{array}{c}
\left< 8 || \overline{15}_{I=\f{1}{2}} || 3 \right> \\
\left< 27 || \overline{15}_{I=\f{1}{2}} || 3 \right> \\
\left< 27 || 24_{I=\f{3}{2}} || 3 \right> \\
\left< 27 || \overline{42}_{I=\f{1}{2}} || 3 \right> \\
\left< 27 || \overline{42}_{I=\f{3}{2}} || 3 \right> \\
\end{array} \right),
\end{equation}
and
\begin{equation}
{\cal A}(B^0_s \rightarrow K^0 \eta_1) = (+1) \left< 1 ||
\overline{15}_{I=\f{1}{2}} || 3 \right> .
\end{equation}
\subsubsection{\rm{$\Delta S = -2$ ($\Delta I_3 = +\f{3}{2}$, $\Delta Y =
-\f{7}{3}$)}}
\begin{equation}
{\cal A} (B^0_d \rightarrow \bar K^0 \bar K^0) = (+1) \left< 27 ||
\overline{42}_{I=\f{3}{2}} || 3 \right> .
\end{equation}
\subsubsection{\rm{$\Delta S = +3$ ($\Delta I_3 = -1$, $\Delta Y =
+\f{8}{3}$)}}
\begin{equation}
{\cal A} (B^0_s \rightarrow K^0 K^0) = (+1) \left< 27 ||
\overline{42}_{I=1} || 3 \right> .
\end{equation}
\subsection{$B \rightarrow D,P$}
\subsubsection{\rm{$\Delta S = 0$ ($\Delta I_3 = 0$, $\Delta Y =
-\f{2}{3}$)}}
For $P$ an octet meson, denote ${\bf u}_{D,8}^{S=0} = {\cal
O}_{D,8}^{S=0} \, {\bf v}_{D,8}^{S=0}$, where
\begin{equation}
{\bf u}_{D,8}^{S=0} =
\left( \begin{array}{c}
{\cal A} (B^+_u \rightarrow D^+ \pi^0) \\
{\cal A} (B^+_u \rightarrow D^+ \eta_8) \\
{\cal A} (B^+_u \rightarrow D^0 \pi^+) \\
{\cal A} (B^+_u \rightarrow D_s^+ \bar K^0) \\
{\cal A} (B^0_d \rightarrow D^+ \pi^-) \\
{\cal A} (B^0_d \rightarrow D^0 \pi^0) \\
{\cal A} (B^0_d \rightarrow D^0 \eta_8) \\
{\cal A} (B^0_d \rightarrow D_s^+ K^-) \\
{\cal A} (B^0_s \rightarrow D^+ K^-) \\
{\cal A} (B^0_s \rightarrow D^0 \bar K^0) \\
\end{array} \right) , \hspace{5em}
{\bf v}_{D,8}^{S=0} =
\left( \begin{array}{c}
\left< \bar 3 || 3_{I=0} || 3 \right> \\
\left< \bar 3 || \bar 6_{I=1} || 3 \right> \\
\left< 6 || 3_{I=0} || 3 \right> \\
\left< 6 || 15_{I=0} || 3 \right> \\
\left< 6 || 15_{I=1} || 3 \right> \\
\left< \overline{15} || \bar 6_{I=1} || 3 \right> \\
\left< \overline{15} || 15_{I=0} || 3 \right> \\
\left< \overline{15} || 15_{I=1} || 3 \right> \\
\left< \overline{15} || \overline{24}_{I=1} || 3 \right> \\
\left< \overline{15} || \overline{24}_{I=2} || 3 \right> \\
\end{array} \right),
\end{equation}
and ${\cal O}_{D,8}^{S=0} =$
\begin{equation}
\left( \begin{array}{cccccccccc}
-\f{1}{4}\s{\f{3}{2}} & -\f{1}{4}\s{\f{3}{2}} & -\f{1}{4\s{2}} &
-\f{1}{4\s{2}} & -\f{1}{4} & -\f{1}{4}\s{\f{5}{2}} & +\f{1}{4\s{6}} &
-\f{\s{3}}{4} & 0 & -\f{1}{\s{3}} \\
+\f{1}{4\s{2}} & +\f{1}{4\s{2}} & -\f{1}{4}\s{\f{3}{2}} &
-\f{1}{4}\s{\f{3}{2}} & -\f{\s{3}}{4} & -\f{1}{4}\s{\f{3}{10}} &
-\f{3}{4\s{2}} & +\f{1}{4} & -\f{1}{\s{5}} & 0 \\
+\f{\s{3}}{4} & +\f{\s{3}}{4} & +\f{1}{4} & +\f{1}{4} & +\f{1}{2\s{2}}
& -\f{3}{4\s{5}} & -\f{1}{4\s{3}} & -\f{1}{2\s{6}} & -\f{1}{\s{30}} &
-\f{1}{\s{6}} \\
-\f{\s{3}}{4} & -\f{\s{3}}{4} & +\f{1}{4} & +\f{1}{4} & +\f{1}{2\s{2}}
& -\f{1}{4\s{5}} & -\f{\s{3}}{4} & +\f{1}{2\s{6}} & -\s{\f{2}{15}} & 0
\\
-\f{\s{3}}{4} & +\f{\s{3}}{4} & -\f{1}{4} & -\f{1}{4} & +\f{1}{2\s{2}}
& -\f{3}{4\s{5}} & +\f{1}{4\s{3}} & -\f{1}{2\s{6}} & -\f{1}{\s{30}} &
+\f{1}{\s{6}} \\
+\f{1}{4}\s{\f{3}{2}} & -\f{1}{4}\s{\f{3}{2}} & +\f{1}{4\s{2}} &
+\f{1}{4\s{2}} & -\f{1}{4} & -\f{1}{4}\s{\f{5}{2}} & -\f{1}{4\s{6}} &
-\f{\s{3}}{4} & 0 & +\f{1}{\s{3}} \\
+\f{1}{4\s{2}} & -\f{1}{4\s{2}} & -\f{1}{4}\s{\f{3}{2}} &
-\f{1}{4}\s{\f{3}{2}} & +\f{\s{3}}{4} & +\f{1}{4}\s{\f{3}{10}} &
-\f{3}{4\s{2}} & -\f{1}{4} & +\f{1}{\s{5}} & 0 \\
-\f{\s{3}}{4} & +\f{\s{3}}{4} & +\f{1}{4} & +\f{1}{4} & -\f{1}{2\s{2}}
& +\f{1}{4\s{5}} & -\f{\s{3}}{4} & -\f{1}{2\s{6}} & +\s{\f{2}{15}} & 0
\\
0 & 0 & -\f{1}{2} & +\f{1}{2} & 0 & -\f{1}{\s{5}} & 0 & +\f{1}{\s{6}}
& +\s{\f{2}{15}} & 0 \\
0 & 0 & -\f{1}{2} & +\f{1}{2} & 0 & +\f{1}{\s{5}} & 0 & -\f{1}{\s{6}}
& -\s{\f{2}{15}} & 0
\end{array} \right),
\end{equation}
and
\begin{equation}
\left(
\begin{array}{c}
{\cal A} (B^+_u \rightarrow D^+ \eta_1) \\
{\cal A} (B^0_d \rightarrow D^0 \eta_1)
\end{array}
\right) = \left(
\begin{array}{cc}
-\f{1}{\s{2}} & -\f{1}{\s{2}} \\
-\f{1}{\s{2}} & +\f{1}{\s{2}}
\end{array} \right)
\left( \begin{array}{c}
\left< \bar 3 || 3_{I=0} || 3 \right> \\
\left< \bar 3 || \bar 6_{I=1} || 3 \right>
\end{array} \right) .
\end{equation}
\subsubsection{\rm{$\Delta S = +1$ ($\Delta I_3 = -\f{1}{2}$,
$\Delta Y = +\f{1}{3}$)}}
For $P$ an octet meson, denote ${\bf u}_{D,8}^{S=1} = {\cal
O}_{D,8}^{S=1} \, {\bf v}_{D,8}^{S=1}$, where
\begin{equation}
{\bf u}_{D,8}^{S=1} =
\left( \begin{array}{c}
{\cal A} (B^+_u \rightarrow D^+ K^0) \\
{\cal A} (B^+_u \rightarrow D^0 K^+) \\
{\cal A} (B^+_u \rightarrow D_s^+ \pi^0) \\
{\cal A} (B^+_u \rightarrow D_s^+ \eta_8) \\
{\cal A} (B^0_d \rightarrow D^0 K^0) \\
{\cal A} (B^0_d \rightarrow D_s^+ \pi^-) \\
{\cal A} (B^0_s \rightarrow D^+ \pi^-) \\
{\cal A} (B^0_s \rightarrow D^0 \pi^0) \\
{\cal A} (B^0_s \rightarrow D^0 \eta_8) \\
{\cal A} (B^0_s \rightarrow D_s^+ K^-) \\
\end{array} \right) , \hspace{5em}
{\bf v}_{D,8}^{S=1} =
\left( \begin{array}{c}
\left< \bar 3 || 3_{I=\f{1}{2}} || 3 \right> \\
\left< \bar 3 || \bar 6_{I=\f{1}{2}} || 3 \right> \\
\left< 6 || 3_{I=\f{1}{2}} || 3 \right> \\
\left< 6 || 15_{I=\f{1}{2}} || 3 \right> \\
\left< 6 || 15_{I=\f{3}{2}} || 3 \right> \\
\left< \overline{15} || \bar 6_{I=\f{1}{2}} || 3 \right> \\
\left< \overline{15} || 15_{I=\f{1}{2}} || 3 \right> \\
\left< \overline{15} || 15_{I=\f{3}{2}} || 3 \right> \\
\left< \overline{15} || \overline{24}_{I=\f{1}{2}} || 3 \right> \\
\left< \overline{15} || \overline{24}_{I=\f{3}{2}} || 3 \right> \\
\end{array} \right),
\end{equation}
and ${\cal O}_{D,8}^{S=1} =$
\begin{equation}
\left( \begin{array}{cccccccccc}
+\f{\s{3}}{4} & -\f{\s{3}}{4} & -\f{1}{4} & -\f{1}{4\s{3}} &
-\f{1}{\s{6}} & -\f{1}{4\s{5}} & -\f{5}{12} & +\f{1}{3\s{2}} &
+\f{1}{3\s{5}} & -\f{1}{3} \\
-\f{\s{3}}{4} & +\f{\s{3}}{4} & -\f{1}{4} & -\f{1}{4\s{3}} &
-\f{1}{\s{6}} & -\f{3}{4\s{5}} & +\f{1}{12} & +\f{1}{3\s{2}} &
-\f{2}{3\s{5}} & -\f{1}{3} \\
0 & 0 & +\f{1}{2\s{2}} & +\f{1}{2\s{6}} & +\f{1}{\s{3}} &
-\f{1}{\s{10}} & -\f{1}{3\s{2}} & +\f{1}{3} & -\f{1}{3\s{10}} &
-\f{\s{2}}{3} \\
+\f{1}{2\s{2}} & -\f{1}{2\s{2}} & 0 & 0 & 0 & -\f{1}{2}\s{\f{3}{10}} &
+\f{1}{2}\s{\f{3}{2}} & 0 & -\s{\f{3}{10}} & 0 \\
0 & 0 & -\f{1}{2} & -\f{1}{2\s{3}} & +\f{1}{\s{6}} & -\f{1}{\s{5}} &
-\f{1}{3} & -\f{1}{3\s{2}} & -\f{1}{3\s{5}} & +\f{1}{3} \\
0 & 0 & +\f{1}{2} & +\f{1}{2\s{3}} & -\f{1}{\s{6}} & -\f{1}{\s{5}} &
-\f{1}{3} & -\f{1}{3\s{2}} & -\f{1}{3\s{5}} & +\f{1}{3} \\
+\f{\s{3}}{4} & +\f{\s{3}}{4} & -\f{1}{4} & +\f{\s{3}}{4} &
0 & +\f{1}{4\s{5}} & -\f{1}{12} & +\f{\s{2}}{3} &
-\f{1}{3\s{5}} & +\f{1}{3} \\
-\f{1}{4}\s{\f{3}{2}} & -\f{1}{4}\s{\f{3}{2}} & +\f{1}{4\s{2}} &
-\f{1}{4}\s{\f{3}{2}} & 0 & -\f{1}{4\s{10}} & +\f{1}{12\s{2}} &
+\f{2}{3} & +\f{1}{3\s{10}} & +\f{\s{2}}{3} \\
-\f{1}{4\s{2}} & -\f{1}{4\s{2}} & -\f{1}{4}\s{\f{3}{2}} &
+\f{3}{4\s{2}} & 0 & -\f{3}{4}\s{\f{3}{10}} & +\f{1}{4}\s{\f{3}{2}} &
0 & +\s{\f{3}{10}} & 0 \\
+\f{\s{3}}{4} & +\f{\s{3}}{4} & +\f{1}{4} & -\f{\s{3}}{4} &
0 & -\f{3}{4\s{5}} & +\f{1}{4} & 0 & +\f{1}{\s{5}} & 0
\end{array} \right),
\end{equation}
and
\begin{equation}
\left(
\begin{array}{c}
{\cal A} (B^+_u \rightarrow D_s^+ \eta_1) \\
{\cal A} (B^0_s \rightarrow D^0 \eta_1)
\end{array}
\right) = \left(
\begin{array}{cc}
+\f{1}{\s{2}} & -\f{1}{\s{2}} \\
+\f{1}{\s{2}} & +\f{1}{\s{2}}
\end{array} \right)
\left( \begin{array}{c}
\left< \bar 3 || 3_{I=\f{1}{2}} || 3 \right> \\
\left< \bar 3 || \bar 6_{I=\f{1}{2}} || 3 \right>
\end{array} \right) .
\end{equation}
\subsubsection{\rm{$\Delta S = -1$ ($\Delta I_3 = +\f{1}{2}$,
$\Delta Y = -\f{5}{3}$)}}
\begin{equation}
\left(
\begin{array}{c}
{\cal A} (B^+_u \rightarrow D^+ \bar K^0) \\
{\cal A} (B^0_d \rightarrow D^+ K^-) \\
{\cal A} (B^0_d \rightarrow D^0 \bar K^0)
\end{array}
\right) = \left(
\begin{array}{ccc}
0 & -\s{\f{2}{3}} & -\f{1}{\s{3}} \\
+\f{1}{\s{2}} & -\f{1}{\s{6}} & +\f{1}{\s{3}} \\
+\f{1}{\s{2}} & +\f{1}{\s{6}} & -\f{1}{\s{3}}
\end{array} \right)
\left( \begin{array}{c}
\left< 6 || 15_{I=\f{1}{2}} || 3 \right> \\
\left< \overline{15} || 15_{I=\f{1}{2}} || 3 \right> \\
\left< \overline{15} || \overline{24}_{I=\f{3}{2}} || 3 \right>
\end{array} \right) .
\end{equation}
\subsubsection{\rm{$\Delta S = +2$ ($\Delta I_3 = -1$,
$\Delta Y = +\f{4}{3}$)}}
\begin{equation}
\left(
\begin{array}{c}
{\cal A} (B^+_u \rightarrow D_s^+ K^0) \\
{\cal A} (B^0_s \rightarrow D^0 K^0) \\
{\cal A} (B^0_s \rightarrow D_s^+ \pi^-)
\end{array}
\right) = \left(
\begin{array}{ccc}
0 & +\s{\f{2}{3}} & -\f{1}{\s{3}} \\
+\f{1}{\s{2}} & +\f{1}{\s{6}} & +\f{1}{\s{3}} \\
-\f{1}{\s{2}} & +\f{1}{\s{6}} & +\f{1}{\s{3}}
\end{array} \right)
\left( \begin{array}{c}
\left< 6 || 15_{I=1} || 3 \right> \\
\left< \overline{15} || 15_{I=1} || 3 \right> \\
\left< \overline{15} || \overline{24}_{I=1} || 3 \right>
\end{array} \right) .
\end{equation}
\subsection{$B \rightarrow \bar D, P$}
\subsubsection{\rm{$\Delta S = 0$ ($\Delta I_3 = +1$, $\Delta Y =
0$)}}
For $P$ an octet meson, denote ${\bf u}_{\bar D,8}^{S=0} = {\cal
O}_{\bar D,8}^{S=0} \, {\bf v}_{\bar D,8}^{S=0}$, where
\begin{equation}
{\bf u}_{\bar D,8}^{S=0} =
\left( \begin{array}{c}
{\cal A} (B^+_u \rightarrow \bar D^0 \pi^+) \\
{\cal A} (B^0_d \rightarrow D^- \pi^+) \\
{\cal A} (B^0_d \rightarrow \bar D^0 \pi^0) \\
{\cal A} (B^0_d \rightarrow \bar D^0 \eta_8) \\
{\cal A} (B^0_d \rightarrow D_s^- K^+) \\
{\cal A} (B^0_s \rightarrow \bar D^0 \bar K^0) \\
{\cal A} (B^0_s \rightarrow D_s^- \pi^+)
\end{array} \right) , \hspace{5em}
{\bf v}_{\bar D,8}^{S=0} =
\left( \begin{array}{c}
\left< 3 || 8_{I=1} || 3 \right> \\
\left< \bar 6 || 8_{I=1} || 3 \right> \\
\left< \bar 6 || \overline{10}_{I=1} || 3 \right> \\
\left< 15 || 8_{I=1} || 3 \right> \\
\left< 15 || 10_{I=1} || 3 \right> \\
\left< 15 || 27_{I=1} || 3 \right> \\
\left< 15 || 27_{I=2} || 3 \right> \\
\end{array} \right),
\end{equation}
and ${\cal O}_{\bar D,8}^{S=0} =$
\begin{equation}
\left( \begin{array}{ccccccc}
0 & 0 & 0 & -2\s{\f{2}{15}} & -\f{1}{\s{6}} & -\f{1}{2\s{5}} &
-\f{1}{2} \\
-\f{1}{2}\s{\f{3}{2}} & +\f{1}{2\s{3}} & -\f{1}{\s{6}} &
-\f{1}{2}\s{\f{3}{10}} & 0 & -\f{1}{2\s{5}} & + \f{1}{2} \\
+\f{\s{3}}{4} & -\f{1}{2\s{6}} & +\f{1}{2\s{3}} &
-\f{1}{4}\s{\f{5}{3}} & -\f{1}{2\s{3}} & 0 & +\f{1}{\s{2}} \\
+\f{1}{4} & +\f{1}{2\s{2}} & -\f{1}{2} & +\f{1}{4\s{5}} & -\f{1}{2} &
+\s{\f{3}{10}} & 0 \\
-\f{1}{2}\s{\f{3}{2}} & -\f{1}{2\s{3}} & +\f{1}{\s{6}} &
+\f{1}{2\s{30}} & -\f{1}{\s{6}} & +\f{1}{\s{5}} & 0 \\
0 & -\f{1}{\s{3}} & -\f{1}{\s{6}} & -\s{\f{2}{15}} & +\f{1}{\s{6}} &
+\f{1}{\s{5}} & 0 \\
0 & +\f{1}{\s{3}} & +\f{1}{\s{6}} & -\s{\f{2}{15}} & +\f{1}{\s{6}} &
+\f{1}{\s{5}} & 0 
\end{array} \right) ,
\end{equation}
and
\begin{equation}
{\cal A} (B^0_d \rightarrow \bar D^0 \eta_1) = (+1) \left< 3 || 8_{I=1}
|| 3 \right> .
\end{equation}
\subsubsection{\rm{$\Delta S = +1$ ($\Delta I_3 = +\f{1}{2}$, $\Delta
Y = +1$)}}
For $P$ an octet meson, denote ${\bf u}_{\bar D,8}^{S=1} = {\cal
O}_{\bar D,8}^{S=1} \, {\bf v}_{\bar D,8}^{S=1}$, where
\begin{equation} \label{bdps1}
{\bf u}_{\bar D,8}^{S=1} =
\left( \begin{array}{c}
{\cal A} (B^+_u \rightarrow \bar D^0 K^+) \\
{\cal A} (B^0_d \rightarrow D^- K^+) \\
{\cal A} (B^0_d \rightarrow \bar D^0 K^0) \\
{\cal A} (B^0_s \rightarrow D^- \pi^+) \\
{\cal A} (B^0_s \rightarrow \bar D^0 \pi^0) \\
{\cal A} (B^0_s \rightarrow \bar D^0 \eta_8) \\
{\cal A} (B^0_s \rightarrow D_s^- K^+)
\end{array} \right) , \hspace{5em}
{\bf v}_{\bar D,8}^{S=1} =
\left( \begin{array}{c}
\left< 3 || 8_{I=\f{1}{2}} || 3 \right> \\
\left< \bar 6 || 8_{I=\f{1}{2}} || 3 \right> \\
\left< \bar 6 || \overline{10}_{I=\f{1}{2}} || 3 \right> \\
\left< 15 || 8_{I=\f{1}{2}} || 3 \right> \\
\left< 15 || 10_{I=\f{3}{2}} || 3 \right> \\
\left< 15 || 27_{I=\f{1}{2}} || 3 \right> \\
\left< 15 || 27_{I=\f{3}{2}} || 3 \right> \\
\end{array} \right),
\end{equation}
and ${\cal O}_{\bar D,8}^{S=1} =$
\begin{equation}
\left( \begin{array}{ccccccc}
0 & 0 & 0 & -2\s{\f{2}{15}} & +\f{1}{\s{6}} & -\s{\f{2}{15}} &
-\f{1}{\s{6}} \\
0 & +\f{1}{\s{3}} & -\f{1}{\s{6}} & -\s{\f{2}{15}} & -\f{1}{\s{6}} &
-\f{1}{\s{30}} & + \f{1}{\s{6}} \\
0 & -\f{1}{\s{3}} & +\f{1}{\s{6}} & -\s{\f{2}{15}} & -\f{1}{\s{6}} &
-\f{1}{\s{30}} & + \f{1}{\s{6}} \\
-\f{1}{2}\s{\f{3}{2}} & -\f{1}{2\s{3}} & -\f{1}{\s{6}} &
+\f{1}{2\s{30}} & +\f{1}{\s{6}} & -\f{1}{\s{30}} & +\f{1}{\s{6}} \\
+\f{\s{3}}{4} & +\f{1}{2\s{6}} & +\f{1}{2\s{3}} & -\f{1}{4\s{15}} &
+\f{1}{\s{3}} & +\f{1}{2\s{15}} & +\f{1}{\s{3}} \\
+\f{1}{4} & -\f{1}{2\s{2}} & -\f{1}{2} & -\f{3}{4\s{5}} & 0
& +\f{3}{2\s{5}} & 0 \\
-\f{1}{2}\s{\f{3}{2}} & +\f{1}{2\s{3}} & +\f{1}{\s{6}} &
-\f{1}{2}\s{\f{3}{10}} & 0 & +\s{\f{3}{10}} & 0
\end{array} \right) ,
\end{equation}
and
\begin{equation}
{\cal A} (B_s^0 \rightarrow \bar D^0 \eta_1) = (+1) \left< 3 ||
8_{I=\f{1}{2}}|| 3 \right> .
\end{equation}
\subsubsection{\rm{$\Delta S = -1$ ($\Delta I_3 = +\f{3}{2}$, $\Delta
Y = -1$)}}
\begin{equation}
\left(
\begin{array}{c}
{\cal A} (B^0_d \rightarrow \bar D^0 \bar K^0) \\
{\cal A} (B^0_d \rightarrow D_s^- \pi^+)
\end{array}
\right) = \left(
\begin{array}{cc}
-\f{1}{\s{2}} & +\f{1}{\s{2}} \\
+\f{1}{\s{2}} & +\f{1}{\s{2}}
\end{array} \right)
\left( \begin{array}{c}
\left< \bar 6 || \overline{10}_{I=\f{3}{2}} || 3 \right> \\
\left< 15 || 27_{I=\f{3}{2}} || 3 \right>
\end{array} \right) .
\end{equation}
\subsubsection{\rm{$\Delta S = +2$ ($\Delta I_3 = 0$, $\Delta
Y = +2$)}}
\begin{equation}
\left(
\begin{array}{c}
{\cal A} (B^0_s \rightarrow D^- K^+) \\
{\cal A} (B^0_s \rightarrow \bar D^0 K^0)
\end{array}
\right) = \left(
\begin{array}{cc}
-\f{1}{\s{2}} & +\f{1}{\s{2}} \\
+\f{1}{\s{2}} & +\f{1}{\s{2}}
\end{array} \right)
\left( \begin{array}{c}
\left< \bar 6 || \overline{10}_{I=0} || 3 \right> \\
\left< 15 || 27_{I=1} || 3 \right>
\end{array} \right) .
\end{equation}
\subsection{$B \rightarrow D, \bar D$}
\subsubsection{\rm{$\Delta S = 0$ ($\Delta I_3 = +\f{1}{2}$, $\Delta
Y = -\f{1}{3}$)}}
\begin{equation}
\left(
\begin{array}{c}
{\cal A} (B^+_u \rightarrow D^+ \bar D^0) \\
{\cal A} (B^0_d \rightarrow D^+ D^-) \\
{\cal A} (B^0_d \rightarrow D^0 \bar D^0) \\
{\cal A} (B^0_d \rightarrow D_s^+ D_s^-) \\
{\cal A} (B^0_s \rightarrow D^+ D_s^-)
\end{array}
\right) = \left(
\begin{array}{ccccc}
0 & -\f{1}{2}\s{\f{3}{2}} & -\f{1}{2} & -\f{1}{2\s{6}} & -\f{1}{\s{3}}
\\
+\f{1}{\s{3}} & -\f{1}{\s{6}} & 0 & -\f{1}{\s{6}} & +\f{1}{\s{3}} \\
-\f{1}{\s{3}} & -\f{1}{2\s{6}} & -\f{1}{2} & +\f{1}{2\s{6}} &
+\f{1}{\s{3}} \\
+\f{1}{\s{3}} & +\f{1}{2\s{6}} & -\f{1}{2} & +\f{1}{2}\s{\f{3}{2}} & 0
\\
0 & -\f{1}{2}\s{\f{3}{2}} & +\f{1}{2} & +\f{1}{2}\s{\f{3}{2}} & 0
\\
\end{array} \right)
\left( \begin{array}{c}
\left< 1 || \bar 3_{I=\f{1}{2}} || 3 \right> \\
\left< 8 || \bar 3_{I=\f{1}{2}} || 3 \right> \\
\left< 8 || 6_{I=\f{1}{2}} || 3 \right> \\
\left< 8 || \overline{15}_{I=\f{1}{2}} || 3 \right> \\
\left< 8 || \overline{15}_{I=\f{3}{2}} || 3 \right>
\end{array} \right) .
\end{equation}
\subsubsection{\rm{$\Delta S = +1$ ($\Delta I_3 = 0$, $\Delta
Y = +\f{2}{3}$)}}
\begin{equation} \label{bdds1}
\left(
\begin{array}{c}
{\cal A} (B^+_u \rightarrow D_s^+ \bar D^0) \\
{\cal A} (B^0_d \rightarrow D_s^+ D^-) \\
{\cal A} (B^0_s \rightarrow D^+ D^-) \\
{\cal A} (B^0_s \rightarrow D^0 \bar D^0) \\
{\cal A} (B^0_s \rightarrow D_s^+ D_s^-)
\end{array}
\right) = \left(
\begin{array}{ccccc}
0 & -\f{1}{2}\s{\f{3}{2}} & +\f{1}{2} & -\f{1}{2\s{2}} & -\f{1}{2}
\\
0 & -\f{1}{2}\s{\f{3}{2}} & - \f{1}{2} & -\f{1}{2\s{2}} & +\f{1}{2}
\\
+\f{1}{\s{3}} & +\f{1}{2\s{6}} & +\f{1}{2} & -\f{1}{2\s{2}} &
+\f{1}{2} \\
-\f{1}{\s{3}} & -\f{1}{2\s{6}} & +\f{1}{2} & +\f{1}{2\s{2}} &
+\f{1}{2} \\
+\f{1}{\s{3}} & -\f{1}{\s{6}} & 0 & +\f{1}{\s{2}} & 0
\\
\end{array} \right)
\left( \begin{array}{c}
\left< 1 || \bar 3_{I=0} || 3 \right> \\
\left< 8 || \bar 3_{I=0} || 3 \right> \\
\left< 8 || 6_{I=1} || 3 \right> \\
\left< 8 || \overline{15}_{I=0} || 3 \right> \\
\left< 8 || \overline{15}_{I=1} || 3 \right>
\end{array} \right) .
\end{equation}
\subsubsection{\rm{$\Delta S = -1$ ($\Delta I_3 = +1$, $\Delta
Y = -\f{4}{3}$)}}
\begin{equation}
{\cal A} (B^0_d \rightarrow D^+ D_s^-) = (+1) \left< 8 ||
\overline{15}_{I=1} || 3 \right> .
\end{equation}
\subsubsection{\rm{$\Delta S = +2$ ($\Delta I_3 = -\f{1}{2}$, $\Delta
Y = +\f{5}{3}$)}}
\begin{equation}
{\cal A} (B_s^0 \rightarrow D_s^+ D^-) = (+1) \left< 8 ||
\overline{15}_{I=\f{1}{2}} || 3 \right> .
\end{equation}
\subsection{$B \rightarrow \eta_c , P$}
\subsubsection{\rm{$\Delta S = 0$ ($\Delta I_3 = +\f{1}{2}$, $\Delta
Y = -\f{1}{3}$)}}
\begin{equation}
\left(
\begin{array}{c}
{\cal A} (B^+_u \rightarrow \eta_c \pi^+) \\
{\cal A} (B^0_d \rightarrow \eta_c \pi^0) \\
{\cal A} (B^0_d \rightarrow \eta_c \eta_8) \\
{\cal A} (B^0_s \rightarrow \eta_c \bar K^0) \\
\end{array}
\right) = \left(
\begin{array}{cccc}
-\f{1}{2}\s{\f{3}{2}} & -\f{1}{2} & -\f{1}{2\s{6}} & -\f{1}{\s{3}} \\
-\f{\s{3}}{4} & -\f{1}{2\s{2}} & -\f{1}{4\s{3}} & +\s{\f{2}{3}} \\
+\f{1}{4} & -\f{1}{2}\s{\f{3}{2}} & +\f{3}{4} & 0 \\
-\f{1}{2}\s{\f{3}{2}} & +\f{1}{2} & +\f{1}{2}\s{\f{3}{2}} & 0
\end{array} \right)
\left( \begin{array}{c}
\left< 8 || \bar 3_{I=\f{1}{2}} || 3 \right> \\
\left< 8 || 6_{I=\f{1}{2}} || 3 \right> \\
\left< 8 || \overline{15}_{I=\f{1}{2}} || 3 \right> \\
\left< 8 || \overline{15}_{I=\f{3}{2}} || 3 \right>
\end{array} \right) ,
\end{equation}
and
\begin{equation}
{\cal A} (B^0_d \rightarrow \eta_c \eta_1) = (+1) \left< 1 || \bar
3_{I=\f{1}{2}} || 3 \right> .
\end{equation}
\subsubsection{\rm{$\Delta S = +1$ ($\Delta I_3 = 0$, $\Delta
Y = +\f{2}{3}$)}}
\begin{equation}
\left(
\begin{array}{c}
{\cal A} (B^+_u \rightarrow \eta_c K^+) \\
{\cal A} (B^0_d \rightarrow \eta_c K^0) \\
{\cal A} (B^0_s \rightarrow \eta_c \pi^0) \\
{\cal A} (B^0_s \rightarrow \eta_c \eta_8) \\
\end{array}
\right) = \left(
\begin{array}{cccc} \label{etac}
-\f{1}{2}\s{\f{3}{2}} & +\f{1}{2} & -\f{1}{2\s{2}} & -\f{1}{2} \\
-\f{1}{2}\s{\f{3}{2}} & -\f{1}{2} & -\f{1}{2\s{2}} & +\f{1}{2} \\
0 & +\f{1}{\s{2}} & 0 & +\f{1}{\s{2}} \\
-\f{1}{2} & 0 & +\f{\s{3}}{2} & 0 
\end{array} \right)
\left( \begin{array}{c}
\left< 8 || \bar 3_{I=0} || 3 \right> \\
\left< 8 || 6_{I=1} || 3 \right> \\
\left< 8 || \overline{15}_{I=0} || 3 \right> \\
\left< 8 || \overline{15}_{I=1} || 3 \right>
\end{array} \right) ,
\end{equation}
and
\begin{equation}
{\cal A} (B^0_s \rightarrow \eta_c \eta_1) = (+1) \left< 1 || \bar
3_{I=0} || 3 \right> .
\end{equation}
\subsubsection{\rm{$\Delta S = -1$ ($\Delta I_3 = +1$, $\Delta Y =
-\f{4}{3}$)}}
\begin{equation}
{\cal A} (B^0_d \rightarrow \eta_c \bar K^0) = (+1) \left< 8 ||
\overline{15}_{I=1} || 3 \right> .
\end{equation}
\subsubsection{\rm{$\Delta S = +2$ ($\Delta I_3 = -\f{1}{2}$, $\Delta
Y = +\f{5}{3}$)}}
\begin{equation}
{\cal A} (B^0_s \rightarrow \eta_c K^0) = (+1) \left< 8 ||
\overline{15}_{I=\f{1}{2}} || 3 \right> .
\end{equation}
\end{document}